# Contributions from Cognitive Research to Mathematics and Science Education[*]

an invited talk given at the

Workshop on Research in Science and Mathematics Education
January 20–24, 1992
Winterton, South Africa

by

**William J Gerace**

*Department of Physics*

*University of Massachusetts*

*Amherst, MA 01003*

[In D. Grayson (Ed.), **Proceedings of the Workshop on Research in Science and Mathematics Education** (pp. 25-44). Pietermaritzburg, South Africa: Teeanem Printers (Pty) Ltd. (1992).]

[*]Work supported in part by NSF Grant # MDR–9050213

*[Note: During the talk on which this paper is based, there were many questions, comments, and observations by the Workshop participants. These have been paraphrased, and appear in italics to distinguish them from the speaker's replies.]*

It has been requested that we make these presentations as interactive as possible. That is fine by me and is much more my style, so please feel free to interrupt me with any questions you may have.

By way of being interactive, I would like to start with a quiz:

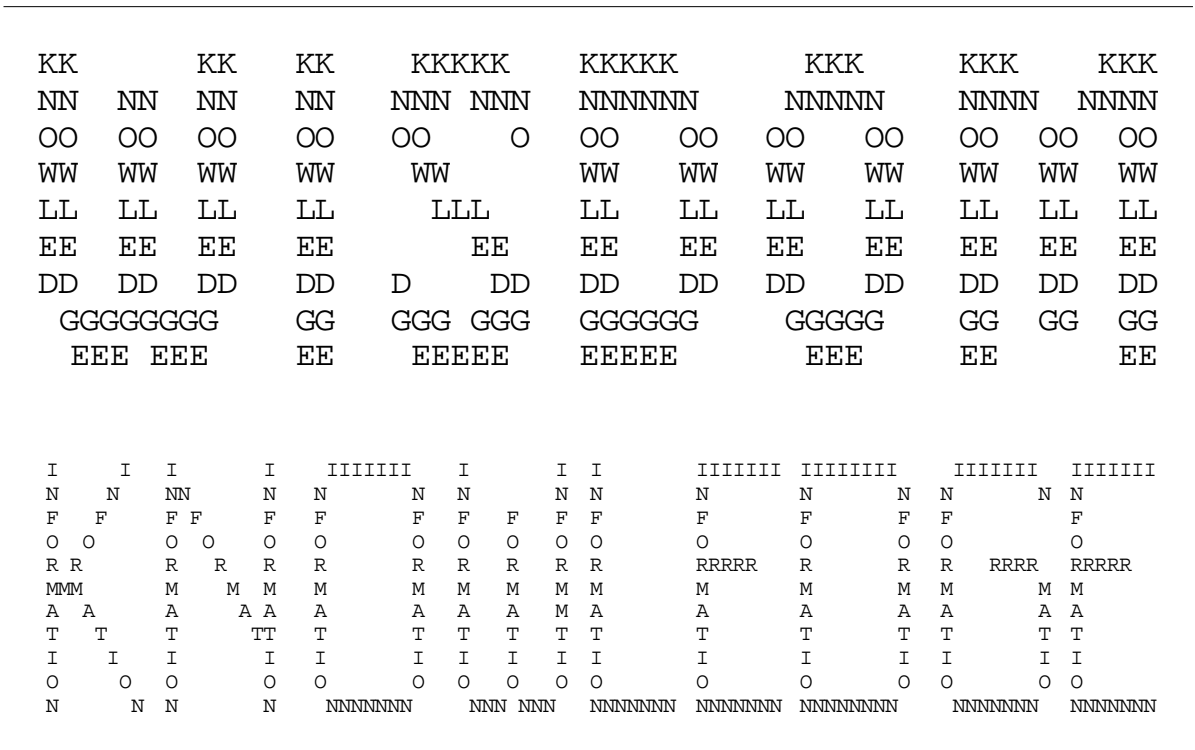

**Fig. 1.** "What do *you* see in this picture?"

What does this figure mean? Just this: from far away you can make out the important words: *wisdom* in the first instance, and *knowledge* in the second. If you are too close you can't perceive the big picture but rather focus on the small, individual letters used to form the big message. The small letters are used to spell *knowledge* in the first case, and *information* in the second. This is parallel to the forest-and-trees analogy.

My apologies to Douglas Hofstadter, who is the author of a book entitled **Gödel, Escher, Bach** [1] where I got the idea for this figure. Perhaps some of you have read it. I believe that it is a valuable book to read and recommend that you try it.

What I have tried to do in this figure is present a representation of what I consider to be the fundamental problem at the moment with mathematics and science education. It is perhaps best explained by a quote from T. S. Eliot's ***Choruses from the Rock***:

> *Where is the wisdom we have lost in knowledge?*
> *Where is the knowledge we have lost in information?*

The point here is that we in science and maths education have allowed ourselves to become information dispensers. We should resist that trend and restore the goal of imparting wisdom.

This talk is divided into four broad areas. These are:

(1)  General remarks about cognitive science
(2)  Hard evidence from cognitive research
(3)  Some descriptive cognitive models
(4)  Contributions from cognitive science to maths and science education.

## General remarks about cognitive science

I should, perhaps, point out at the outset that the views expressed in this talk are my own; I make no claim that they are authoritative or even represent the consensus view of cognitive researchers.

I will begin my general remarks by addressing the motivation for doing cognitive science. As far as I am concerned, a major motivation is what I alluded to in the introductory remarks. This might be defined as *information pollution*. As in other endeavors, it is time for us in science and maths education to take a more "ecological" perspective. We are being inundated with a lot of facts, and the prevailing view is that students ought to know these facts. I think that is the wrong way to go about it. What you want to do is treat your students as thinkers. Teach them to teach themselves. Make them learners, and then you don't have to worry about teaching them all the facts; they'll do that themselves.

What is the aim of cognitive science? It is to achieve an understanding of the processes underlying knowledge acquisition and utilization. We devise models to help explain the processes of learning and problem solving. For us in the field of pedagogy, research on cognitive processes offers the potential for developing educational practices that are more efficient and/or applicable to a wider audience.

Who are the practitioners or contributors to cognitive research?  This is very important because, in my opinion, you, as educators, should be playing a significant role.  There is a group of people who call themselves cognitive scientists, but those people are generally in psychology departments and tend to be concerned about things like eye movements and neurons in the brain.  Cognitive science is a very broad-based spectrum, ranging all the way from that kind of psychology right on up to pedagogy.  You (as educators) cannot afford to wait for some specialist or some expert to come along and tell you how you should teach or what you should teach.  You should be engaged in discovering those answers yourselves.  You are as competent as anybody to make those discoveries, and people are making them all the time.  One thing that is clear to me is that cognitive science is not going to bear much fruit until content people and educators get very strongly involved in the research process.

What is the nature of cognitive science?  At present, it is very qualitative.  All we have are general, qualitative descriptions of certain kinds of processes.  Precise or quantitative models are not yet possible.  Artificial Intelligence (AI) would like to have such quantitative and procedural models of learning because these models could then be used to make a computer learn.  AI is a major driving force towards that direction of research.  For the near future, cognitive models will remain descriptive.

What is the methodology of cognitive science?  In the previous talk, David Treagust described various types of research studies and their associated methodologies.  That is not what I'm talking about.  The methodology I am referring to here is of a more broad or fundamental nature and this I would characterize as a *modified* scientific paradigm.  What is the scientific paradigm?  Just simply: observe, form models, and then perform experiments to verify the models.  This is what all scientists do.  A modified paradigm is needed for cognitive science because there is a large question here about what constitutes observation; in particular, we have to extend observation to include reflection: observation *of ourselves*; reflections on our own learning processes.

In this vein, I would like to comment on something that worries me about cognitive science in general, and that is whether or not the mental structures and processes that we have for the *performance* of our activities as scientists are adequate for the task of *evaluating* our activities.  A visual representation of what this means is shown below.

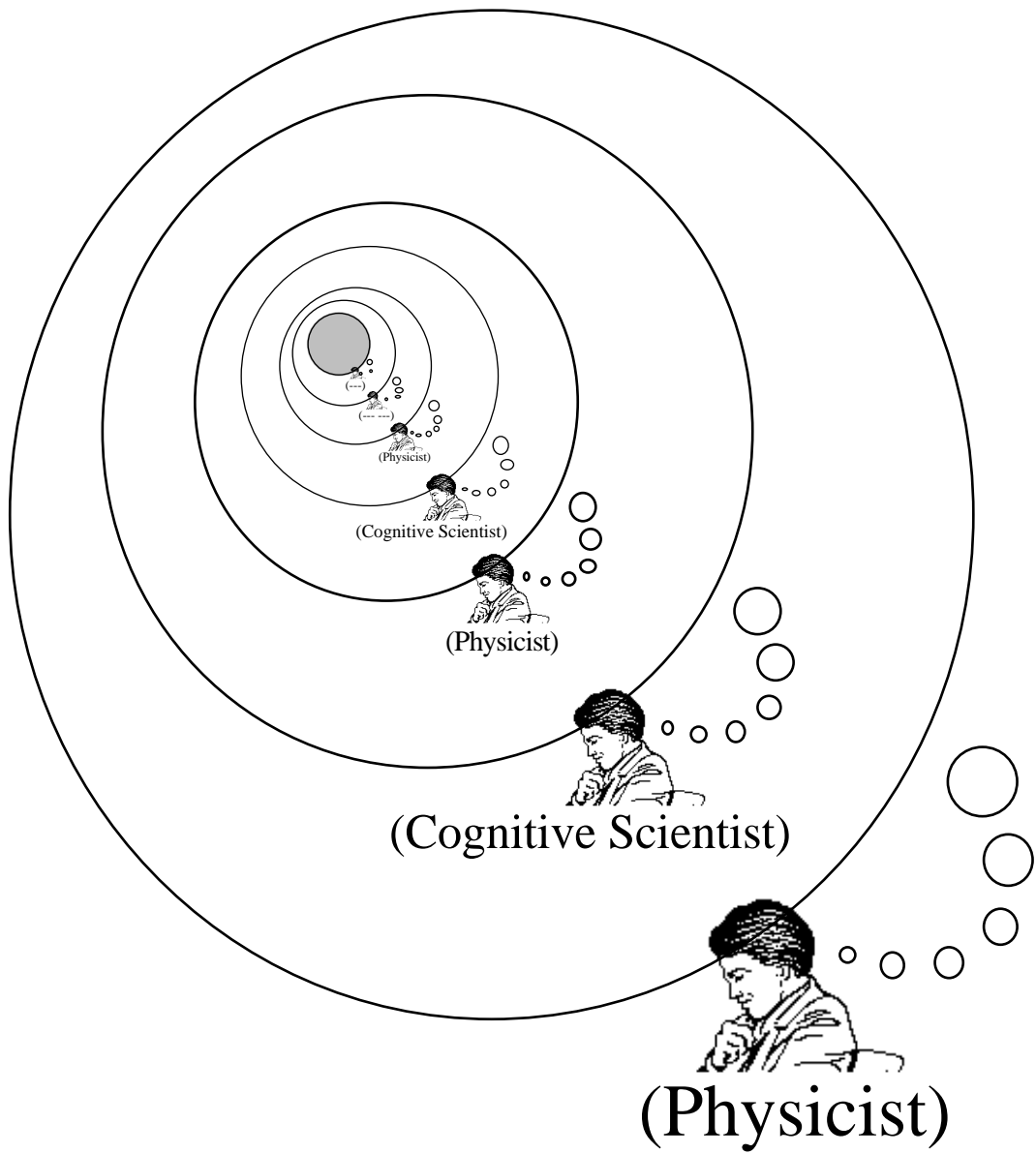

**Fig. 2.** A physicist thinking about himself as a cognitive scientist thinking about himself as a physicist …

Any physicists in the audience will recognize this as a self-energy diagram. The diagram would apply equally well to any science or mathematics educator. You ponder yourself or you ponder the thoughts and activities of a cognitive scientist (which you should also be), and you should be thinking about how you think. And how you think you think. And how you think you think you think. Etc.

A few examples might make these points about thinking about thinking and the insufficiency of our cognitive processes more clear. When I teach physics, I try to make students aware of the fact that they already have a lot of information but do not have the wherewithal to extract the information from their experience. This information is, therefore, not available to them and they cannot use it. I think a similar situation occurs with teachers like yourselves. You have a lot of information—an awful lot of experience dealing with your interactions with students. If you could extract this information, and communicate it to the rest of us, we all would benefit.

Extracting information from your memory store presumes the ability to interact willfully with your recall processes. Think of some piece of information that you haven't used in an long time; maybe it was a phone number when you were in college or the name of somebody you met ages ago. Why haven't you forgotten that? Can you forget it? Clearly there are aspects of information acquisition, storage, and deletion that are not voluntary at all. You cannot willfully make yourself forget something and thereby free up that memory space for something else. Fortunately the human brain never seems to be full, so I don't think you should worry about that limitation.

What does it mean to know something? How do you know that you know it? I invite you to consider certain distinctions that we all make. Sometimes you know that you know something. Sometimes you know that you knew something, but you don't know it any longer. Sometimes, you sit there and say to yourself, "I think I know that, but I don't know where it is." Such pondering brings to mind a paraphrase of the famous Descartes quote: "Cognito ergo sum" ("I think, therefore I am.") that I once saw scrawled on a bathroom wall at MIT. It was: "I think I think, therefore I am… I think."

*Do you distinguish between knowing and understanding?*
Yes; but there is a deep inter-relationship between the two. Understanding means that you can use the knowledge as opposed to merely have it. And that requires, from my point of view, that the knowledge be integrated into some kind of a mental structure.

*So it's possible to know without understanding?*
Oh yes; I know lots of things I don't understand.

*Does that refer to something like an address or something that you recall?*
Understanding implies to me that the information is useful. What does it mean to use a certain kind of information? An address might be useful for locating some other place. If you don't know the address, you can't use it. You may know it and still not be able to use it if you can't place it in a context; if you don't know where it is in a city, you can't orient yourself by that location.

My final general remarks about cognitive science concern constructivism. First of all, I believe that constructivism is fundamental to cognitive science. Why? Well, to me, constructivism plays the same role for cognitive science that causality plays for physics. It's a fundamental viewpoint; it doesn't answer any question for you. All it does is affect how you ask the questions and the kinds of answers you'll accept, but it doesn't itself provide any answers. From a philosophical point of view, I think that constructivism is almost self-evident. Further, there is no disharmony between constructivism and a rational or scientific perspective.

I would caution you, however, that the term constructivism, as used in the literature, has a broad spectrum of meanings. The first task you must do is decide which one of those meanings the author intends. The term constructivism may be used: (1) to indicate a philosophy; (2) to indicate a theory of epistemology; (3) to indicate a theory of communication; or (4) to imply a pedagogic approach. Unfortunately, the mapping from constructivism to a pedagogic approach is non-unique. As a matter of fact, you'll find violent disagreements among people about what constitutes a constructivist approach to education.

The implications of constructivism for communication and pedagogy will be considered in the last section of this talk. As a philosophy, constructivism places severe limits upon our ability to know anything about, or even to determine the existence of, an external reality. These restrictions derive from the premises of constructivism as an epistemology and it is those premises that concern us here. They are:

- Knowledge is constructed, not transmitted.
- Prior knowledge impacts the learning process.
- The construction of knowledge requires purposeful and effortful activity.
- Initial understanding is local, not global.

These assertions will be explored in more detail later after we have discussed some cognitive models. Such models provide a context in which it is easier to understand the operational aspects of the premises. The only point I want to make here is that the constructivist point of view is clearly divergent from earlier views of education that presumed we could put information into a student's head. From a constructivist perspective, real learning can occur only when the learner is actively engaged in operating on, or mentally processing, the incoming stimuli. Furthermore, the interpretation of those stimuli depends upon previously constructed learning. Such processing may or may not be conscious. Depending upon the degree of self-consciousness, meta-cognitive issues may be present.

The fundamental moral for pedagogy is simply this: when you send a message to a student (by saying something or providing information), if you have no knowledge of the receiver, then you have no idea as to what message is received, and you cannot interpret the student's response. You do not know what the student is doing with that information or what learning, if any, occurs.

*Could you explain the difference between Knowledge and Information?*
Not really. It's a very fuzzy issue. We often use the terms interchangeably. I don't think that there is much value to be gained by trying to draw a rigid distinction between the two. We should keep in mind, though, that there are different levels or categories; a piece of information, such as an address, is not the same kind of knowledge as a physics or mathematics concept.

*Couldn't you distinguish them by distinguishing between knowledge of specifics and knowledge of principles?*
A perfectly good partition. Any partition that makes sense is fine.

*Wouldn't it be more appropriate to side-step the matter since all of our understanding of words is contextually based? Since we all must construct our own knowledge of what is meant by Knowledge, it is unlikely that we will all agree to specific definitions or discriminations.*
You certainly may.

*Would you define what you mean by local vs. global understanding before you move on?*
Certainly. As a student forms a concept, even when they can repeat that concept back to you and appear to understand, their understanding is in a very limited context. Outside that context it is as though you never taught them at all; they don't have the concept, or they appear to lose the concept. Consider this example: you teach students to identify the power of a variable, such as $x^2$ or $x^3$, as the little number to the upper right of the $x$. The expression $\sin^2 x$ constitutes a sufficiently different context that they, in some sense, have "lost" the concept.

*Instead of worrying about categories of Knowledge and Information, why not just define the kind of learning in which you are interested? Perhaps we should not concern ourselves with how an address is learned, but only with what it means to understand physics or mathematics. In this way one can avoid introducing constructivism or epistemology. You can simply say, "Here is an approach which is suitable for a specific kind of learning."*

I think it is certainly possible to take such a minimalist or pedagogically pragmatic approach, but cognitive science seeks a more general description of the learning process.

## Hard evidence from cognitive research

The amount of hard evidence gleaned to date from cognitive research studies on high-level science and mathematics learning is underwhelming.  One should keep in mind, however, that the research is very difficult and that the field is in its infancy.  Such studies have tended to concentrate upon the presence and consequences of misconceptions and novice–expert differences.

The topic of mis- or pre-conceptions has proven to be a very fertile area of research.  Here one is usually interested in an erroneous world-view or conceptual framework that is based upon experiences and is formed prior to any formal instruction.  From a constructivist point of view we should not be surprised that, given the proclivity of the human mind to seek patterns, individuals will naturally form a conceptual framework in an effort to account for, and cope with, their experiences.  Important preconceptions have been found in many knowledge domains.  Among these are physics, mathematics, biology, astronomy, and others [2].

Prior conceptual frameworks have been shown, in some instances, to impede learning of a more correct or formal framework.  I am unaware of any specific study that suggests the cause of this, but I do not think the result surprising.  Again, it is a tenet of constructivism that the existing conceptual framework will serve as a filter of subsequent experience.  For example, presented with a physics demonstration of some principle, individuals are more likely to focus their attention on those aspects or features of the phenomenon that are in accord with their current framework.  As a result, accumulating sufficient evidence to disabuse an individual of their previous framework can be quite difficult.

Student misconceptions have been shown to be very difficult to uproot.  They are retained even after a concerted effort has been made to dislodge them.  This is quite natural and we should not be dismayed by it.  It should be remembered that, even apart from the student's emotional attachment to their own viewpoint, they are being asked to abandon ideas that have served a very important cognitive function for them.  Most individuals, including ourselves, would rather retain a half-baked generalization together with a list of exceptions to the rule rather than rethink the entire morass.  Only after this list of exceptions grows unmanageable or the existing generalization has been demonstrated to be inadequate for some important situation will anyone be induced to seek a better generalization.  Herein lies, I believe, the key to attacking student misconceptions.  The more we know about their individual conceptual framework, the better able we are to manifest its flaws and, thereby, induce the student to reformulate their world-view.

Except for the fact that prior experience plays a significant, or perhaps dominant, role in their formation, very little is known about the detailed origins of misconceptions. Not being well informed in this area, I do not have much to say. I do find some of the results quite intriguing. It has been suggested that prior experience and the resulting conceptual framework seem to hinder cognitive development in mathematics less than in science [3]. These findings could conceivably be due to the pervasiveness of mathematics, at least in our western culture. Such a pervasiveness would afford lots of opportunities for discovering inadequacies in self-formed conceptual frameworks. Finally, even if substantiated, such findings could well be culture dependent.

The other area of concentration of cognitive research into higher-order learning is that of novice–expert differences [4]. These studies have revealed a variety of findings regarding the way that experts differ from novices in the nature, structure, and utilization of their knowledge store. Many of these conclusions are indirect, being deduced from behavior patterns while problem solving.

Memory-recall studies have shown that the superior recall capability of experts is intimately related to the employment of high-order knowledge structures. Such recall studies have been made in the fields of chess, electronic circuitry, and computer programming. The methodology used in all of these studies is essentially the same, so only the electronic-circuitry study will be discussed. Both experts and novices at electronics are presented with a circuit diagram and, after only a brief time interval during which to study the diagram, they are asked to recall as much of the diagram as they can. The result is that experts have a significantly superior ability to recall the circuit when the circuit represents a realistic situation, but their advantage is diminished, or even disappears, when the circuit is comprised of randomly arranged electronic elements. This phenomenon has been labeled *chunking* and is believed to be due to the expert's ability to perceive relationships between the individual elements and recognize the function served by clusters of elements. Viewing the circuit diagram, the expert chunks the information—that is, subsumes clusters of individual elements into a single complex entity—thereby greatly facilitating recall. Obviously, this ability vanishes when the logical substructure is absent as it is in the diagrams consisting of randomly arranged elements. The interpretation is that experts have a richly interconnected knowledge store and these interconnections serve an important function for the acquisition of further knowledge. It also follows that establishing these interconnections should be a priority for the developing novice.

*It would seem plausible to me that experts would remember more, even for the random circuits, just because they are experts.*
That doesn't seem to be the case. Once your short-term memory is filled, it is filled and that is all there is to it.

*How strong is the transfer of these processes? In other words, how transferable is that process to other things that the electronics expert might be asked to remember?*
In response to that I have two remarks. The first is that I am unaware of any evidence that would indicate that expertise transfers across domains. Since the implication here is that the expert is analyzing the circuit at a higher level than the detailed level available to the novice and that they're seeing content or relationships that the novice does not see, it seems unlikely that an economics expert, for example, would be able to perceive those, or any other, relationships between elements and is, therefore, for all intents and purposes, equivalent to a novice. My second comment is this. Although expertise in one domain does not directly transfer to another insofar as knowledge structures or problem solving is concerned, I think that an expert may well have a learning advantage in a new domain because of the expert's proclivity to seek relationships. This opinion is based upon observed behavior patterns of good students, which is a topic that will be considered later in this talk.

In the domain of physics, another interesting difference between the expert's and novice's knowledge store is revealed by "categorization" type judgment tasks. These studies show that, in addition to being richly interconnected, the expert's knowledge store is arranged or, more precisely, activated in a hierarchical manner [4]. This is one of the areas in which our group (at UMass) has been working. I must confess to being uninformed about any similar investigations in other domains.

These categorization tasks consist of asking experts and novices to make judgments or decisions about some set of problems. In the original study, subjects were given a set of problems and told to sort, or classify, the problems according to the similarity of their solutions. The problems were of an elementary nature, such as typically found in an introductory course, and the subjects did not solve the problems. The objective was to determine what the subjects perceived to be important about the problems prior to finding solutions.

Results indicate that novices tend to classify problems by the superficial aspects of the problem. For example, they might place all problems involving an inclined plane together and all the problems involving a pendulum together, etc. Experts, on the other hand, tend to sort the problems by the physics principles that could be employed to solve the problem. The tendency of the novice to focus on the "surface features" of the problem, as opposed to the "deep structure" perceived by experts, has been interpreted as evidence that the expert's knowledge store has an organizational structure that is lacking in the novice's.

A more detailed picture has emerged from observations of novices and experts engaged in the problem-solving process. Presented with a problem, experts generally first perform a qualitative analysis of the problem employing principles and concepts. Only after completing this stage does the expert employ relevant operational knowledge—the procedures, relationships, and equations appropriate to the problem situation. The algebraic manipulations necessary to generate a quantitative solution occur at the very final stage of problem solving. For purposes of problem solving, then, it appears that the expert's knowledge has an essentially hierarchical structure, with conceptual knowledge having precedence over operational knowledge. An example of a hierarchical structure for the domain of classical mechanics is shown below in Figure 3 [5].

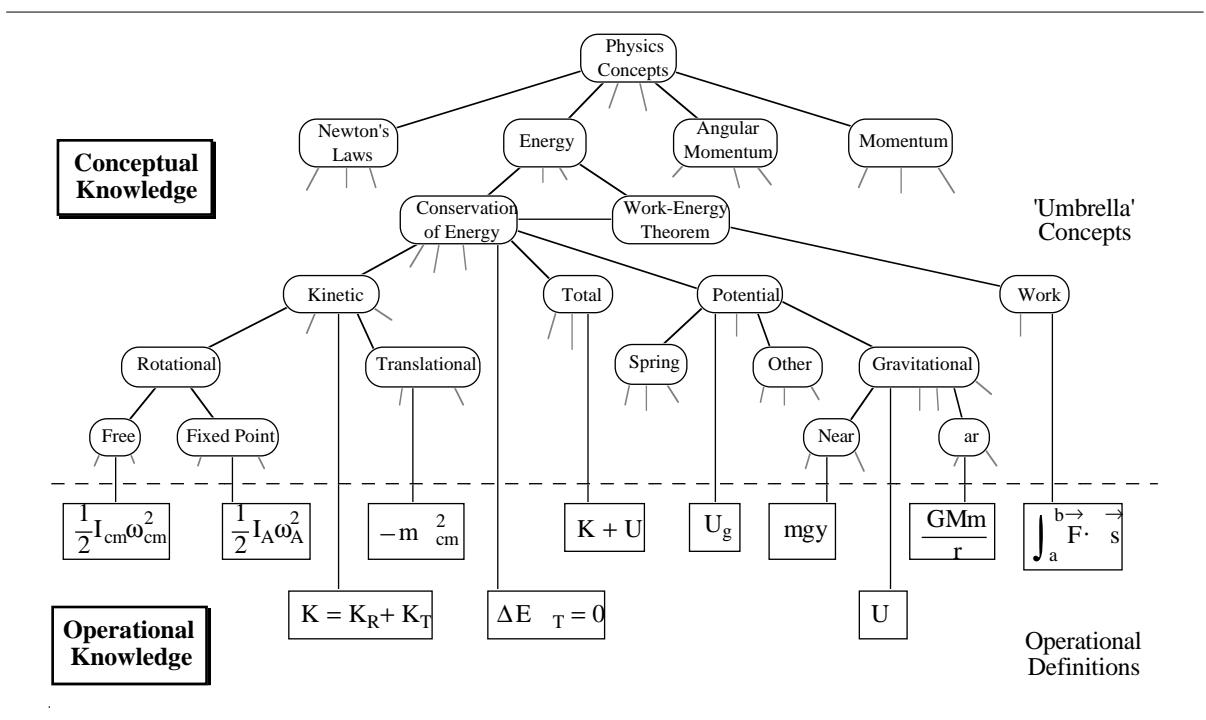

**Fig. 3.** An expert's knowledge structure for classical mechanics [5]

In contrast to this, novices perform what is called a *means–ends analysis*. They focus primarily upon equations and immediately attempt to manipulate them to isolate the desired quantity, often inserting numerical values from the very beginning of the process. Novices do not perceive the strategic value of conceptual analysis, and are apparently distracted by the immediate objective of obtaining an answer.

One of the research projects currently being worked on by our group is an effort to determine the types of activities that are useful for inducing novices to structure their knowledge in a more expert-like fashion [6]. In the original study, a sample of novices were constrained by a computer-based program to perform an expert-like conceptual analysis of problems prior to solving them. The results showed that a student's initial approach to problems could be influenced relatively quickly. After only a rather brief exposure to this device, which incidentally was completely devoid of any instructional feedback, students displayed significant shifts in their tendency to categorize problems by principle rather than by surface feature.

At present we are engaged in a study to determine the pedagogic value of requiring students to provide a qualitative analysis before solving a problem. The requirement is imposed by grading this analysis in exams and in homework. The study involves a large-enrollment introductory mechanics course. Results are not yet final but look promising, at least to the extent of promoting the importance of the fundamental principles.

There are two other research results that I think are very important [3,7]. These are based upon studies of student learning and, while direct comparisons to experts are not made, the findings relate strongly to the expert–novice work. The first result is that good or successful students, who might be regarded as semi-expert, are always seeking patterns and, in general, tend to notice, expand, and refine implicit knowledge [3]. These students are more likely to perceive commonalities of approach to problems or the importance of certain concepts, even when these are not made explicit.

The other finding from student-learning research is that traditional problem-solving activities impose a severe cognitive load upon students [7]. By traditional activities I mean problems that ask students to obtain answers. This emphasis upon the goal of answers predisposes students to adopt a means–ends problem-solving approach, which is often an efficacious way to proceed. The point here is not that a means–ends approach is bad. Experts frequently will resort to such an approach in the absence of some other schema they can apply to the problem situation. The point is that many students can be so intent upon the short-term goal of achieving an answer that they have few cognitive resources to devote to long-term objectives, such as noticing patterns or implicit knowledge. Assorted pedagogical morals can be gleaned from all of these research results and these will be addressed in the last portion of this talk.

*If knowledge is not transferable, which is one of the premises of constructivism, what is the point of knowing how an expert stores knowledge or solves problems, since you cannot transfer that knowledge to the novice?*

You use your knowledge of how an expert thinks to govern your interaction with the student to assist them in constructing a similar structure. As a matter of fact, one of the fundamental points here is that constructivism and cognitive science very much impact one's view of what education means. It is no longer a matter of sending out messages and dispensing information. Teaching is a very interactive process, and requires a lot of bi-directional communication. In his talk, David Treagust gave an excellent example of this need by showing us that many students were able to answer a particular chemistry question correctly without being able to give the correct reason. Simply giving problems and getting answers back is insufficient, even, or perhaps especially, when the answer is correct. If you are to determine what learning actually took place, you need to know why they chose that answer.

Although my comments so far hardly exhaust the topic of novice–expert research, we must move on. Some of the important differences between the knowledge store and the problem-solving behavior of experts and novices have been summarized in Table 1, shown below. The meaning of some of the entries will become clearer as we progress.

**Table 1.** A summary of expert–novice differences

|  | **Experts** | **Novices** |
|---|---|---|
| **Knowledge Characteristics** | Large store of domain-specific knowledge | Sparse knowledge set |
|  | Knowledge richly interconnected and hierarchically structured | Disconnected and amorphous structure |
|  | Integrated multiple representations | Poorly formed and unrelated representations |
| **Problem-Solving Behavior** | Conceptual knowledge impacts problem solving | Problem solving largely independent of concepts |
|  | Performs qualitative analysis | Manipulates equations |
|  | Uses forward-looking concept-based strategies | Uses backward-looking means–ends techniques |

## Some descriptive cognitive models

The cognitive models that will be discussed are macroscopic models. They are called descriptive because they attempt to provide a qualitative description of either categories of knowledge or some cognitive process. Quantitative models, if ever possible, are not likely in the near future. Fully microscopic models, which explain thinking in terms of neuron activation, may never be possible. In any case we should not be deterred or discouraged. There are many examples of situations where great progress has been made despite the absence of complete understanding of the underlying processes. Medicine, for example, has made tremendous progress while a full understanding of cell biology has yet to be achieved.

It is important to remember that, by their very nature, these descriptive models are imperfect. Do not expect them to be free of fault or above criticism. When evaluating such models we must keep in mind their function—the purpose for which they were created. Some models are useful because they provide guidance when designing instructional materials or practices. Other models attempt to explain a complex process in terms of simpler stages. In general, all models are an attempt to increase our understanding of something complicated by decomposing it into elements that are more primitive or fundamental.

One attempt to describe the expert's facility at problem solving decomposes the expert's knowledge and experience relevant for the task of problem solving into four areas [3]. These are: (1) domain knowledge; (2) repertoire of problem-solving skills; (3) meta-cognitive processes; and (4) meta-level knowledge.

The expert is able to draw upon extensive domain-specific knowledge that is stored and accessed as *schema* [7]. The storage of knowledge as schema means that stored along with the knowledge is a body of ancillary facts and experience that permits the expert to recognize situations for which the knowledge is useful. Also stored as part of a schema is a set of rules or procedures for applying the knowledge in a useful way. These associated factors are likely to play a significant role in the process of structuring knowledge hierarchically.

Contained within the repertoire of problem-solving skills are a set of strategies or procedures that the expert might apply to a specific problem situation. Occupying a prominent position in this list would be the rule to activate a schema that has been recognized to have been effective in the past. Other rules might be: transform the problem to an analogous one and see if that is recognized; translate the problem into a different representation; or (when all else fails) use means–ends analysis.

Other factors that are very important for successful problem solving but are difficult to describe are relegated to the area of meta-cognitive processes. Notable among these factors are motivation and self-confidence, both of which have a task-related as well as a task-independent component.  Also included here are a variety of self-monitoring processes that are used to allocate cognitive resources or evaluate the quality of cognitive functioning.

The last category, that of meta-level knowledge, is often neglected but plays a crucial role in the problem-solving process.  One's belief and value systems, which govern all of behavior, are to be found in this area.  Also residing in this region is one's world-view—one's perception of the natural order of things—which can have a profound influence on judgment and behavior.

Needless to say, the expert usually has the edge over the novice in all four areas (domain knowledge, problem-solving skills, meta-cognitive processes, and meta-level knowledge), and certainly has a commanding lead in the first two areas. When helping novices develop, although we tend to concentrate our efforts in the first two areas, we are not unmindful of the impact of these latter areas.  Few would disagree with the assertions that motivation, self-confidence, and reward structure are intricately interwoven or that a student's perception of their ability to solve a problem greatly influences their success rate.  Consciously or not, we consider these factors all the time.  Why, after all, do we arrange exams or tests so that easy problems occur first, and difficult problems appear near the end?  We try to build the student's confidence by giving them some early successes because we know that, if they are discouraged by an inability to do early problems, the exam is apt to be an unreliable measure of the student's ability.

When we counsel students to persevere, we are attempting to interact with their world-view.  But what is their world-view?  Students tend to believe that problems are solved quickly, or not at all.  If, after looking at a problem, they do not see how to solve it in the first two or three minutes, they conclude that the problem is beyond their abilities and give up. Also implicit in their world-view is their presumption that the solution to any problem must use the most recently taught knowledge or technique.  Often we inadvertently strengthen this perception by failing to give comprehensive tests which force the student to integrate their knowledge.  Last, but not least, we should be sensitive to interactions between the student's and our own world-views.  My previous point provides a perfect example.  If we do give a comprehensive test, or an "old" problem, and students do manifest their penchant to apply recent procedures, we are tempted to conclude either that the student lacks intelligence or that they have failed to learn the previous material and therefore we must re-teach it.

*You have stated that one of the features of an expert is that they organize and seek patterns. You have also humorously characterized aspects of the student's world-view, such as trying to*

*do problems with the most recently taught procedure. Isn't the student also organizing all the time and this is just the result of their seeing a pattern in the way that we teach them?* That is very possibly the case. If we changed our teaching methods, that component of the student's world-view might change. On the other hand, the situation might not be so simple. This tendency might be the result of the difficulty involved in making a judgment. It is easier to just continue doing what they have been doing.

Another cognitive model tries to describe the attainment of expertise as a sequence of five developmental stages [8]. Although the divisions between the stages are arbitrary and difficult to delineate, the decomposition is useful for emphasizing several points. The first stage is Novice, and is characterized by someone who is rule dominated. The Novice is aware of some of the rules and their entire perspective is based on rules. Confronted by a situation, they search known rules for an applicable one and, failing that, seek some other, unknown rule for handling the situation. How often have we heard a student say, "I don't know the formula for that case."?

The second stage is that of Advanced Novice. The Advanced Novice has amassed a large set of rules and is experienced at their application. The next stage is labeled Competency. In this stage students remain predominantly rule oriented but the weaning process has begun. A modicum of judgment has crept in but any genuine perspective continues to be absent. Success at this stage fosters self-confidence.

An extremely important stage is the fourth, which is called Proficiency. By this time the student has developed a significant capacity for judgment, and problem-solving behavior is governed by this judgment rather than the application of rules. What makes this stage so important is the onset of self-motivation. Prior to this stage the student is not successful enough to derive much reward from their activities. To derive an internal reward requires an emotional investment. If the risk of frustration or failure is too high, one will not make that investment. To be sure, in the global realm of life, no one, including any one of us, is fully internally motivated; we all retain some external motivators. The issue here, however, is a very limited one. What does it take to become an invested or self-motivated learner? The answer is simple: it requires a reasonable expectation of success. I will return to this point later.

The final stage is that of Expert. The characteristics of the Expert are mature judgment (although perhaps only in their field of expertise), self-motivation, and the ability to self-evaluate. In this model the meta-cognitive aspects of the expert are emphasized, while the knowledge and skill components are assumed.

*Doesn't this model have more to do with motivation than cognition?*
Not necessarily. I fear that I may have given that impression by using the model as an excuse for stressing motivational or psychological factors. Personally, I don't think we can divorce learning from psychology, at least in the later stages of development. I think that we often try to encourage students by giving them simpler problems, or giving them the same problem over again, or something like that. What we should do to help them build confidence and judgment is give them activities that allow them to practice the individual component skills involved in problem solving. In this way they will perceive progress and develop self-motivation. I will be suggesting a few such activities later in the talk when we turn to pedagogic practices.

Another simple descriptive model details the steps involved in the problem-solving process [3]. The first step is to Understand the Problem. During this step the solver encodes the information contained in the problem and forms an internal representation of the problem. Clearly, the encoding and representing processes depend strongly upon the knowledge store. The second step, that of Devising a Plan, will be facilitated by the hierarchical structure of the knowledge and will likely utilize one or more stored schemata. Operational and procedural knowledge plays the major role during the third step, which is Executing the Plan. The last and, perhaps, the most important step in problem solving is Reflection and Evaluation. This is where meta-cognitive factors are brought to bear. This is also usually the hardest step for novices to master.

The three models mentioned above are attempts to describe the ingredients, development, and functioning of the expert. In and of themselves, they do not contribute much new information. Their value lies primarily in their ability to organize information and to enable us to discuss what is a very complicated subject.

The last model I want to discuss is a model that concentrates less on the problem-solving characteristics of the expert, and more on the general structure of knowledge and how that structure is used differently by experts and novices to solve problems. This is our model—the one we use to govern our own curriculum-development work. We find it helpful for thinking about knowledge structures and for discussing the classification of and relationships between knowledge elements. Since our curriculum-development materials and the model on which they are based is the topic of a workshop to be presented later today, I will be very brief here.

A visual representation of the model appropriate for an expert is shown in the figure below. Domain knowledge is partitioned into three types: conceptual knowledge, operational/procedural knowledge, and problem situations. As indicated in the figure, the expert's knowledge is highly structured. A corresponding figure for the novice would be devoid of most of the clustering connections and the bi-directional links between the three types of knowledge.

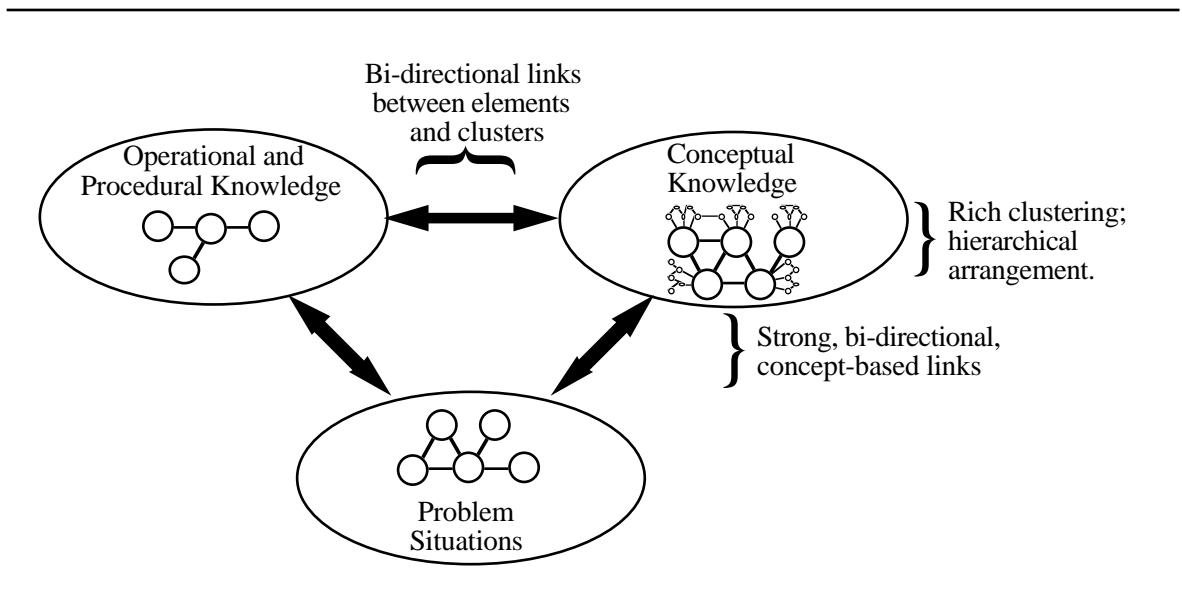

**Fig. 4.** A representation of the expert's structure of knowledge

The only reason for introducing the model here is that I will be alluding to it later when pedagogic practices are discussed. To head off confusion it is, perhaps, beneficial to state explicitly what the model does not contain. Meta-level processes are not treated. Also absent are the procedural aspects of problem solving, although a very similar diagram could likely be devised.

## Cognitive research and maths and science education

What is or ought to be the interaction between cognitive research and maths/science education? At the research level the boundary between these two fields is extremely fuzzy. I think the only difference between cognitive research into high-level thinking and learning, and educational research is fundamentally one of motivation: cognitive researchers would like to understand the underlying processes, while educational researchers are frequently interested in the efficaciousness of some particular pedagogic practice. In any case, the classroom can serve as a valuable laboratory for testing practices derived from cognitive models. As I remarked earlier, I believe both fields would benefit from increased involvement of educators.

The underlying philosophical framework of cognitive science, namely constructivism, stresses the perspective that all learning is an internal process of actively constructing knowledge. The chief implication of this view for education, which is an attempt to interact effectively with the learning process, is that teaching is essentially a communication process. As a consequence, the problem of defining good teaching methods is rephrased as a problem of establishing effective communication channels with the student. Before proceeding to the issue of how we might use communication to assist students in their process of constructing knowledge, I want to discuss the steps involved in forming a reliable channel of communication.

To my way of thinking, the algorithm for evolving a dependable means of communication consists of four steps:

(1) create or refine a "code" or language comprehensible to both the sender and the receiver;
(2) formulate a model of the receiver;
(3) make inferences based upon this model when constructing the message; and
(4) observe some effect or consequence of sending the message.

The last item is an attempt to determine if the message received was the message sent. The observed effect might be a behavior change, a return message, or some other indication that a signal has indeed been received. If the response is as anticipated based upon the model, the process can cycle directly to step (3). In all other instances, the process recycles to step (1) because it is necessary to reconsider the adequacy of either the code or the model of the receiver.

There are two interesting observations to note here. First, these four steps parallel the "handshake" procedures used by two computers or other electronic devices when establishing a communication link. Also intriguing are the associations that can be drawn between these steps and the four steps of expert problem solving mentioned earlier. Viewed an appropriate way, refining the communication process is just another problem to be solved. It is, in fact, the all pervasive problem of the teacher. Simply presenting material, giving students problems, and accepting answers back is not a refined-enough process of communication for efficient education. Failure to solve this problem makes all other efforts ineffectual, if not downright futile.

We turn now to some general pedagogic practices implied by constructivism and by the cognitive research that has been done to date. You will, no doubt, perceive the influence of our own model in the manner in which these generalizations are expressed. Although I will give examples in an effort to be clear, these examples are all drawn from physics. Some additional thought may be needed for you to see specific applications to your own field.

> *1. **Use multiple representations**. A representation may be linguistic, abstract, symbolic, pictorial, or concrete, just to name a few categories. Using many different representations for the same knowledge helps the student to inter-relate knowledge types and relate the knowledge to physical experience.*

We in the sciences are often especially guilty of neglecting to relate the linguistic representation, in which we have originally expressed the concept, to other representations, particularly a concrete one. Relating the symbolic and linguistic representations is also very important. If you wrote an equation on the board, could your student read that equation to you? Conversely, if you were to say to the student, "Write an equation that expresses the fact that the change in the sum of the kinetic energies of two bodies equals the negative of the change in the interaction potential energy between them," what would they write down? If they can't write anything, then all the words you use while solving a problem "just don't compute". Seeing little or no relationship between what you say and what you write, it is little wonder that they focus upon the equations you use for solving the problem. After all, they will be asked to solve problems and their survival depends upon being able to use the equations.

Another example would be subscripts or other subtle distinctions you wish to make. I am confident that all the physicists in the audience have had this experience. You assign a problem involving two masses, labeled m sub 1 ($m_1$) and m sub 2 ($m_2$) or, equivalently, Cap m ($M$) and small m ($m$). By the time the student has written two or three lines of equations the distinction is completely lost. All masses have become $m$, tempting them to cancel the masses out of the problem. Teaching students how to read symbolic expressions will sensitize them to these distinctions.

> *2. **Help students interconnect their knowledge**. These interconnections may be between knowledge elements of the same type, which we label clustering, or between different types, which we call linking. Clustering and linking are facilitated by explicitly identifying similarities and differences between elements.*

As an example of what I mean by explicitly identifying similarities and differences, consider the following case from physics. (My apologies to the non-physicists.) When covering rotational dynamics, I might well say to students, "Torque is a vector quantity. In that way it is like a force, but it does not have the units of force. Torque is measured in the same units as work or energy, which are scalars, but torque is <u>not</u> a type of energy. You should note that units are no longer sufficient to uniquely identify the nature of a quantity."

Creating these clusters and links helps students to weave their knowledge into a fabric, thereby providing them access to the knowledge by means other than chronological. What I mean by that is, in the absence of subsequent processing, we tend to store our knowledge and experience chronologically. You can convince yourself of this by considering what you typically do when you have misplaced something and can't find it. It is like trying to find something on a videotape. You skip backward on your mental tape to find the last point when you remember having the object, then forward again to find the first time you missed the object. You then scrutinize all your actions between these two points. I know, for a fact, that this analogy is applicable to student knowledge. Suppose that a student thinks that they do not know something, but you are convinced that they do, perhaps because they tested well when the material was originally covered. If, by asking them questions, you can get them to recall some other piece of knowledge temporally close to the forgotten information, the missing knowledge often becomes available to them.

This need to create associations that are other than temporal between knowledge elements is also the rationale behind cumulative examinations. Students generally have an immense distaste for such exams. Often I jokingly accuse them of adhering to the displacement theory of learning: in order to learn a new piece of knowledge, they must forget something they knew previously. As long as knowledge is only stored chronologically, this is not far from the truth. Consider how the strategy for locating a lost object becomes impractical if the interval to be searched is a year or more.

> **3. Use extended context to hone concepts**. *Concepts can be very context dependent. They do not become globally useful until they can be abstracted. Exploration of a broad context of applicability helps the student to refine and abstract concepts.*

When presenting new ideas or concepts to students, most teachers try very hard to be precise. However, even when we think we are being excruciatingly lucid, to the student's mind there remains a lot of ambiguity. As students struggle to construct the concept, they look for patterns and generalities. Presented within a very restricted context, they often find many potential generalities and they are confused as to which one should form the kernel of the concept. Further, the better or more attentive the student, the greater the number of possible patterns they perceive. An extended context not only helps the student refine the pattern and, thereby, abstract the concept, but also serves to make both student and teacher aware of remaining latent ambiguities.

An example might help to make this less abstract. In physics we have the concept of a *normal force*, which is the component of the contact force between two objects acting perpendicular to the contact surface. Note first, that the concept has a remaining ambiguity: there are two directions that are perpendicular to any surface. The actual direction of the normal force is assigned at the time that the concept is instantiated or applied to a specific situation and is determined by which of the two objects is being considered.

The non-physicists may be a little befuddled about now. That's OK, you do not need to fully understand normal forces to understand the point I am trying to make. It might even engender some sympathy for the student condition.

My point is actually rather simple. This concept is introduced quite early in high school physics courses. The set of situations in which the concept is used is extremely restricted, often involving only an isolated block resting or moving along a horizontal surface. In such cases the magnitude of the normal force (which is vertical) is equal to the weight of the block. As a result of this limited context, students often conclude that the normal force always acts in the vertical direction and/or that it is always equal to the weight of the block. Such misconstructions can easily be made manifest by placing a second block on top of the first and asking the student to identify the normal force on the bottom block due to the horizontal surface; or by having the block travel along a vertical, curved track. Unchallenged, these erroneous concepts tend to harden and, by the time the student attends college, they can be nearly impossible to correct.

> ***4. Use comparisons and contrasts to sensitize students to categories and relationships.*** *Essential to the process of structuring knowledge is the classification and inter-relation of knowledge elements. Simultaneous consideration of similar or contrasting situations helps students perceive the commonalities and distinctions needed to organize their knowledge store.*

Frequently students fail to perceive any distinction between two objects or situations and, as a result, they classify them the same way; they do not notice the subtle difference(s).

Further, they may fail to see what two apparently very different objects or situations have in common. This lack of sensitivity impedes their ability to form useful categories of and relationships between knowledge elements.

The best way to get students to perceive subtle differences is to present them with two items that differ only in the regard you want them to focus upon. Examining these two items concurrently will reveal the distinction to them. A rather simplistic example might be this: to get students to perceive a subtle shade difference in color, present them with two objects that have slightly different colors but have all other physical attributes in common. If there is a significant difference in the shapes of the items, that becomes the focus of attention, and masks the difference in color.

The best way to get students to perceive some common property between two apparently diverse objects or situations is to enlarge the set being examined. Persisting with our trivial example: someone who is presented with a red ball and a red vase is more likely to notice the commonality of color if they are told that the common feature is shared with another set of objects consisting of a red book, a red chair, and a red flower.

More realistic examples are abundant in physics. Consider some simple equations students encounter in an introductory course: $F=ma$, $F=mg$, $F=-kx$, $F=\mu N$, $F=mv^2/r$. To the student these all appear alike. They do not understand what the fuss is about. For the benefit of the non-physicists, the first equation is the dynamical principle that forms the basis of all of Newtonian mechanics; the next three are approximate empirical laws; and the last is either a special application of the first or (something I personally consider intellectually reprehensible) the definition of a *centripetal force*. The only sense in which all of these equations are alike is their algebraic form. To become an expert physicist, it is the differences that must be understood. Students can be induced to perceive the difference between the first and the middle three by presenting them together and then noting that only the first permits someone to determine the kinematic response of a body subject to forces, thus placing it in a class by itself. The commonality between the other three, namely their empirical nature, can be stressed by noting that they are similar to Ohm's law for circuits and other experimental relationships.

One final comment I would like to make is that I do not believe in what could be called *one-pass learning*. Confronted with student's lack of understanding, teachers are tempted to slow their presentation of the material. This is the wrong way to go and runs the risk of both boring and frustrating students. It is unrealistic to expect students to comprehend and integrate all of the concepts and procedures the first time around. In order for knowledge to be structured into an integrated whole, students must be able to relate knowledge elements. The recognition of relationships between individual knowledge elements requires that some rudimentary form of the knowledge be already present. *Multiple-pass learning*, in which earlier topics are revisited expressly for the purpose of establishing relationships and perspective, is a much better way to go.

> **5. Use more goal-free activities.** *Goal-free activities reduce the cognitive load associated with traditional problem solving. This permits the student to devote a larger amount of their cognitive resources to tasks such as honing concepts or seeking relationships.*

What is meant by a "goal-free activity"? It is this: any question, problem or project that does not have a well-defined or numerical answer. For example, students might be asked to compare two situations or explain how some quantity would change if a given situation were modified slightly. In addition to reducing cognitive load, such activities can be used to reveal erroneous concepts or flaws in the student's knowledge structure. A wrong answer to a traditional problem usually does not permit access to this information. There may be several correct ways to do a problem, but the number of wrong ways is infinite. The incorrect answer, either alone or with intermediate steps included, is insufficient for a teacher to identify uniquely the incorrect path taken by the student. One might go so far as to say that it is testimony to the ingenuity of the human mind that anyone learns anything simply by having their performance graded. Goal-free activities provide a much richer basis for communication.

An example of a goal-free activity in physics would be this: A block is released from rest a distance $d$ from the bottom of a frictionless inclined plane. A small marble and a cylinder, each having the same mass as the block, roll the same distance down similar planes inclined at the same angle. Which of the objects has the largest kinetic energy when it reaches the bottom of the incline, and why do you think this is so? Again, for the benefit of the non-physicist, the three objects all have the same kinetic energy at the bottom. If students were to say that they think the block has the greatest kinetic energy because it is moving the fastest, then they would be telling you either that they misinterpreted the question or that they do not consider rotational motion as having kinetic energy. Subsequent questions (such as, "How do the initial and final energies of each of the objects compare?") would tease apart the various possible misunderstandings.

*How does your suggestion of using more goal-free activities rather than problems reconcile with the efficiency of teaching problem solving?*

I think the issue here is "efficiency at what?"  If the goal is to have students be able to solve some large but finite number of problems, then doing problems all the time may be most efficient.  If, however, the goal is to have students think and be able to solve problems unlike any they have seen before, then using goal-free activities, which help students form schemata, is a better way.

> **6. *Meta-communicate about the learning process*.** *Meta-communication helps students formulate meta-knowledge and increases the likelihood of their becoming self-invested learners.*

The idea of meta-communicating is not new.  Although all teachers already do it to one degree or another, they tend to confine meta-communication to motivational factors.  What is being advocated here is that we should be meta-communicating about the learning process itself.  That, in essence, is what we are doing right now.

As teachers, our objective should not be to train students.  One trains animals; one educates students.  As sentient beings, students can help monitor the communication process.  The chances of having the correct message received is greatly enhanced if there is a conscientious receiver seeking the message.  To my way of thinking, students should be made aware of all the issues we have been discussing.  Their ability to comprehend, of course, will depend upon their innate ability and degree of development.  That just means that we must explain the learning process often.

Meta-communication is a powerful means for helping students to become self-invested learners.  Listed below are some of the active roles that students can play in the learning process, together with an explanation of how meta-communication can be used to help them define and function in that role.

> (a) <u>Students can actively participate in the establishment of a communication language.</u>  Students are capable of appreciating the need for an unambiguous communication language; therefore, forewarning them of the need for precision is very helpful.  This can be reinforced by telling them what a particular term does not mean as well as what it does mean.  In the sciences and mathematics, where terms are rigorously defined, it is especially important to push against colloquial interpretations.  If the term being introduced has a common usage, having students first identify their current meaning for a word helps them to distinguish the new definition from the old.  Examples from physics are terms like "work" or "deceleration". You mathematicians, if you haven't already done so, should ask your students what the term "function" means to them.

(b) <u>Students can be made to be defensive learners.</u> They should be informed of common pitfalls or misunderstandings. I think that this tactic can be very effective in the area of misconceptions, especially for those deep-seated misconceptions that impede further learning. Students can attack common misconceptions easier from within than we can from the outside. While it is not an easy skill for them to learn, students can be shown how to explore deliberately the internal consistency of their mental models. An example from physics would be this: often students think that at the highest point in the trajectory of a projectile the acceleration is zero, usually because the velocity is zero and they do not understand the difference between these two concepts. Asked to specify the velocity and acceleration of an object sitting on a table, they will respond correctly. If they are subsequently asked, "Why then is the behavior of these two objects not the same?" they will often realize their own mistake. The most important part of this interaction is that they have been given an example of how to use their own knowledge to root out errors.

(c) <u>Students can consciously participate in the structuring of knowledge.</u> Students should be told that one of their objectives is to categorize knowledge elements and perceive relationships and patterns. They should be told what it means to solve a problem, and they should be encouraged to notice and extract general procedural patterns. You want to teach the student what it means to meta-think— to be self-conscious about their thinking process. This helps the student form perspective and see the forest as well as the trees. An example of an activity that encourages this would be as follows: given a set of two or more problems, ask the students to solve the problems and then, by comparing the solutions, specify a set of steps or procedures that are common to all the solutions.

I think that often we do not give students enough credit for their ability to participate in the learning process. All of the roles mentioned here sooner or later occur naturally in the good student. Meta-communication can help to speed up the process.

*I think I get the feel of a Catch-22. Starting from a constructivist view, which emphasizes the individual, you make up a set of rules the teacher should use when interacting with the learner. Hasn't the individual again been lost?*

I can understand your concern. However, the rules that emerge are so general that they actually encourage you to adjust the communication process to take account of both your own style as well as that of the individual student. Implicit in the rules is the idea that there is not a perfect set of curriculum materials or a perfect lecture that every student will understand. Another point I would like to make is that the set of steps for evolving a reliable channel of communication apply not only to student–teacher interactions, but also to student–student interactions. By being attentive to the process, teachers can help students improve their communication skills in general, thereby greatly enhancing peer learning. Students do not need the teacher all the time. By interacting and communicating about the subject matter with other students, they can do a lot of refining and constructing of concepts on their own. I think cooperative learning, or learning groups, should play a significant role in modern education.

*I would like to return to your point about students being distracted from real learning by the need to obtain answers. Isn't that our own fault and the fault of our entire education system?*

To a certain extent I would agree. We must keep in mind, however, that, while it is not the only important goal, getting the correct answer is still important. We cannot completely remove the need for answers or tests of problem solving. There is an aspect to the current stress upon answers which may be even worse than distraction. We may have rewarded students so much for getting the right answer that we have locked them into a means–ends approach, which is efficient for getting answers. The more successful they are, the more rewarded, and the harder they are to teach. Students can become addicted to that kind of success, and very fearful of abandoning any process that yields correct answers.

When most students get to college, which is where I tend to see them, they try to carry over thought and study patterns that they used in high school. When these fail them, they panic. I tell them that no one can tell them how they should learn. No two people learn exactly the same way. Some people need to do problems; other people are more visual, they can just read more; some people are very interactive, and learn better that way. They really have to become very experimental learners, try to figure out how they learn best and determine what kinds of experiences they should seek. I think you can begin to make that point to them at a much earlier age than college.

*I'd like to just comment that, in an examination system, the rewards in college are for the same things as in high school: getting answers.*

A very important point. If you want to change the system you have to interact with the assessment process. You have to reward students for thinking, not just getting the right answers. You must be activist, both in terms of the exams you yourself make up, as well as interacting with whatever state assessment processes exist. You have to make conceptual issues more prevalent.

To summarize, I would assert that the greatest contribution of cognitive research to mathematics and science education is a re-formulation of the entire educational endeavor as one that should be:

(1) **Learner-centered.** Methodologies must take account of prior learning. Students should be self-conscious learners and fully engaged in the process.
(2) **Process-oriented.** Knowledge is constructed and not transmitted. Students must actively process their experience to form useful knowledge structures.
(3) **A bi-directional communication process.** Effective education requires two-way interaction with the learner.

In closing I would like to return to the issue I raised at the very beginning. As both scientists and educators, we should reaffirm that Wisdom, and not Knowledge or Information, should be the primary goal of education.

Thank-you…

# References

*[Note: No effort has been made to provide an exhaustive list of references. Interested readers should consult these references and the references contained therein.]*

1. Douglas R Hofstadter, **Gödel, Escher, Bach: an eternal golden braid**, (New York: Basic Books, 1979).

2. Lillian Christie McDermott, *Millikan Lecture 1990: What we teach and what is learned— closing the gap*, **American Journal of Physics 59**(4), 301–315 (April 1990); Mark L Rosenquist & Lillian C McDermott, *A conceptual approach to teaching kinematics*, **American Journal of Physics 55**(5), 407–415 (May 1987).

3. Edward A Silver & Sandra P Marshall, *Mathematical and Scientific Problem Solving: Findings, Issues and Instructional Implications*. In BF Jones & L Idol (Eds.), **Dimensions of thinking and cognitive instruction, Vol. 1**, (Hillsdale, NJ: Lawrence Erlbaum Associates, 1990).

4. Michelene TH Chi, Paul J Feltovich & Robert Glaser, *Categorization and Representation of Physics Problems by Experts and Novices*, **Cognitive Science 5**, 121–152 (1981); Stephen K Reed, Carolyn C Ackinclose & Audrey A Voss, *Selecting analogous problems: Similarity versus inclusiveness*, **Memory & Cognition 18**(1), 83–98 (1990).

5. Figure taken from: Jose P Mestre, *Learning and instruction in pre-college physical science*, **Physics Today 44**(9), 56–62 (September 1991).

6. Jose Mestre, Robert J Dufresne, William Gerace, Pamela Hardiman & Jerold Touger, *Enhancing higher-order thinking skills in physics*. In D Halpern (Ed.), **Development of Thinking Skills in the Sciences and Mathematics**, (Hillsdale, NJ: Lawrence Erlbaum Associates) 77–94 (1992); Jose Mestre, Robert J Dufresne, William J Gerace, Pamela T Hardiman & Jerold S Touger, *Promoting a Greater Reliance on Principle Use Among Beginning Physics Students*, **Journal of Research in Science Teaching**, (accepted for publication); Robert J Dufresne, William Gerace, Pamela Hardiman & Jose Mestre, *Constraining Novices to Perform Expert-like Problem Analyses: Effects on Schema Acquisition*, **Journal of the Learning Sciences**, (accepted for publication).

7. John Sweller, *Cognitive Technology: Some Procedures for Facilitating Learning and Problem Solving in Mathematics and Science*, **Journal of Educational Psychology 81**(4), 457–466 (1989); John Sweller & Graham A Cooper, *The Use of Worked Examples as a Substitute for Problem Solving in Learning Algebra*, **Cognition & Instruction 2**(1), 59–89 (1985); John Sweller, *Cognitive Load During Problem Solving: Effects on Learning*, **Cognitive Science 12**, 257–285 (1988).

8. Hubert L Dreyfus & Stuart E Dreyfus, **Mind over Machine: The Power of Human Intuition and Expertise in the Era of the Computer**, (New York: The Free Press, A Division of Macmillan, Inc.) Chapter 1, *Five Steps from Novice to Expert*, 16–51 (1986).